\documentclass[prl,showpacs,amsmath,amssymb,twocolumn,twoside,superscriptaddress]{revtex4}

\usepackage{dcolumn}
\usepackage{bm}
\usepackage{graphicx}
\usepackage{color}

\begin{document}

\newtheorem{corollary}{Corollary}
\newtheorem{definition}[corollary]{Definition}
\newtheorem{example}[corollary]{Example}
\newtheorem{lemma}[corollary]{Lemma}
\newtheorem{proposition}[corollary]{Proposition}
\newtheorem{theorem}[corollary]{Theorem}
\newtheorem{fact}[corollary]{Fact}
\newtheorem{property}[corollary]{Property}
\newtheorem{observation}[corollary]{Observation}

\def\squareforqed{\hbox{\rlap{$\sqcap$}$\sqcup$}}
\def\qed{\ifmmode\squareforqed\else{\unskip\nobreak\hfil
\penalty50\hskip1em\null\nobreak\hfil\squareforqed
\parfillskip=0pt\finalhyphendemerits=0\endgraf}\fi}
\def\endenv{\ifmmode\;\else{\unskip\nobreak\hfil
\penalty50\hskip1em\null\nobreak\hfil\;
\parfillskip=0pt\finalhyphendemerits=0\endgraf}\fi}
\newenvironment{proof}{\noindent \textbf{{Proof~} }}{\qed}
\newenvironment{remark}{\noindent \textbf{{Remark~}}}{\qed}
\newcommand{\remarkTitle}[1]{\textbf{(#1)}}
\newcommand{\proofComment}[1]{\exampleTitle{#1}}

\newcommand{\bra}[1]{\langle #1|}
\newcommand{\ket}[1]{|#1\rangle}
\newcommand{\braket}[3]{\langle #1|#2|#3\rangle}
\newcommand{\ip}[2]{\langle #1|#2\rangle}
\newcommand{\op}[2]{|#1\rangle\!\langle #2|}

\newcommand{\tr}{{\operatorname{Tr}\,}}
\newcommand{\supp}{{\operatorname{supp}\,}}
\newcommand{\Sch}{{\operatorname{Sch}}}
\newcommand{\GHZ}{{\textrm{GHZ}}}

\newcommand{\rk}{{\operatorname{rk}}}
\newcommand{\sr}{{\operatorname{srk}}}
\newcommand{\pr}{{\operatorname{pr}}}

\newcommand{\E}{{\mathcal{E}}}
\newcommand{\F}{{\mathcal{F}}}
\newcommand{\h}{\mathcal{H}}
\newcommand{\diag} {{\rm diag}}

\newcommand{\nc}{\newcommand}
\nc{\ox}{\otimes}

\def\a{\alpha}
\def\b{\beta}
\def\g{\gamma}
\def\d{\delta}
\def\e{\epsilon}
\def\ve{\varepsilon}
\def\z{\zeta}
\def\t{\theta}
\def\k{\kappa}
\def\l{\lambda}
\def\m{\mu}
\def\n{\nu}
\def\x{\xi}
\def\p{\pi}
\def\r{\rho}
\def\s{\sigma}
\def\ta{\tau}
\def\u{\upsilon}
\def\ph{\varphi}
\def\c{\chi}
\def\ps{\psi}

\def\G{\Gamma}
\def\D{\Delta}
\def\T{\Theta}
\def\L{\Lambda}
\def\X{\Xi}
\def\P{\Pi}
\def\S{\Sigma}
\def\U{\Upsilon}
\def\Ph{\Phi}
\def\Ps{\Psi}
\def\O{\Omega}

\title{Tensor Rank and Stochastic Entanglement Catalysis for Multipartite Pure States}

\author{Lin Chen}
\affiliation{Centre for Quantum Technologies, National University of
Singapore, 3 Science Drive 2, Singapore 117542}
\email{cqtcl@nus.edu.sg}

\author{Eric Chitambar}
\affiliation{Physics Department, University of Michigan, 450 Church Street,
Ann Arbor, Michigan 48109-1040, USA}

\author{Runyao Duan}
\affiliation{Centre for Quantum Computation and Intelligent Systems (QCIS),  Faculty of Engineering \protect\\ and Information Technology, University of Technology, Sydney, NSW 2007, Australia}
\affiliation{State Key Laboratory of Intelligent Technology and Systems, Tsinghua National Laboratory \protect\\ for Information Science  and Technology, Department of Computer Science and Technology, \protect\\ Tsinghua University, Beijing 100084, China}

\author{Zhengfeng Ji}
\affiliation{Perimeter Institute for Theoretical Physics, Waterloo, ON, N2L 2Y5, Canada}

\author{Andreas Winter}
\affiliation{Department of Mathematics, University of Bristol, Bristol BS8 1TW, U.K.}
\affiliation{Centre for Quantum Technologies, National University of
Singapore, 3 Science Drive 2, Singapore 117542}

\date{\today}

\begin{abstract}
The tensor rank (also known as generalized Schmidt rank)  of multipartite pure
states plays an important role in the study of entanglement
classifications and transformations.  We employ powerful tools from
the theory of homogeneous polynomials to investigate the tensor rank
of symmetric states such as the tripartite state
$\ket{W_3}=\tfrac{1}{\sqrt{3}}(\ket{100}+\ket{010}+\ket{001})$ and
its $N$-partite generalization $\ket{W_N}$.  Previous tensor rank
estimates are dramatically improved and we show that (i) three
copies of $\ket{W_3}$ has rank either $15$ or $16$, (ii) two copies
of $\ket{W_N}$ has rank $3N-2$, and (iii) $n$ copies of $\ket{W_N}$
has rank $O(N)$.  A remarkable consequence of these results is that
certain multipartite transformations, impossible even
probabilistically, can become possible when performed in multiple
copy bunches or when assisted by some catalyzing state.  This
effect is impossible for bipartite pure states.
\end{abstract}

\pacs{03.67.-a, 03.65.Ud, 03.67.Hk}

\maketitle

Multipartite entanglement has attracted
increasing attention due to its intriguing properties and potential
applications in both quantum information processing and condensed matter physics~\cite{Wang-2009a, Guhne-2005a}.
A central question in the subject concerns the convertibility between different multipartite
entangled states by using local operations and classical
communications (LOCC).
If such a protocol is only stochastic (i.e., occurs with a non-zero
probability) then we say that
the two states are \emph{convertible via stochastic LOCC (SLOCC)}; when the transformation is reversible, the two states are called \emph{SLOCC equivalent}.  In bipartite systems, SLOCC convertibility is characterized by the Schmidt rank of the state: bipartite $\ket{\psi}$ is SLOCC convertible to $\ket{\phi}$ if and only if the Schmidt rank of $\ket{\psi}$ is no smaller than that of $\ket{\phi}$.

A generalization of the Schmidt rank in multipartite systems and
also relevant to SLOCC transformations is the \emph{tensor rank}.
Formally, for states in $N$-partite quantum systems, each of which
is described by a $d$-dimensional Hilbert space $\h_i$
($i=1,\ldots,N$), the tensor rank $\rk(\psi)$ of a state $\ket{\psi}
\in \bigotimes_{\alpha=1}^N \h_\alpha$, defined as the smallest
number of product states
$\{\bigotimes_{\alpha=1}^N\ket{\phi^\alpha_i}\}_{i=1...rk(\psi)}$
whose linear span contains $\ket{\psi}$.  The tensor rank has been
extensively studied in algebraic complexity theory
\cite{Kruskal-1977a, buergisser-book}, and while it is easy to
compute for $N=2$ (Schmidt rank), even for $N=3$, determining the
rank of a state is NP-hard \cite{Haastad-1990a}.  This is one reason
why SLOCC convertibility in multipartite systems is so challenging.

Despite this general difficulty, the analysis becomes less formidable when certain classes of states are considered such as symmetric states, i.e. those invariant under any permutation of its parties.  Recently, symmetric states have received much attention in the study of entanglement measures \cite{Hubener-2009a, Chen-2010a} and bound entanglement \cite{Toth-2009a}.  Furthermore, the entanglement transformation properties of symmetric states have been investigated \cite{ Bastin-2009b} and experimental procedures have been designed which use symmetric states in generating families of multiqubit SLOCC equivalent states \cite{Bastin-2009a, Wieczorek-2009a}.

There is a natural correspondence between symmetric tensors and
symmetric polynomials where the theory of homogeneous polynomials
can be used to study the latter.  As we will explore in greater
detail, every homogenous polynomial possesses a quantity called the
polynomial rank which is closely related to the tensor rank.  The
relationship between the two ranks allows for known results on the
polynomial rank to be used directly on tensor rank
estimations~\cite{cgl08}.  This method will prove to be quite
powerful.

It is easy to see that the tensor rank is an SLOCC monotone: if $\ket{\psi}$ can be transformed into $\ket{\phi}$ via SLOCC, then $\rk(\psi)\geq \rk(\phi)$.  In general the converse is not true~\cite{DVC00}, however for any state SLOCC equivalent to the $d$-level $N$-partite GHZ state $\ket{GHZ^d_N}=\frac{1}{\sqrt{d}}\sum_{i=1}^d\ket{i}^{\ox N}$, tensor rank does decide convertibility~\cite{CDS08}.
\begin{observation}
  \label{le:ghztype}
  A GHZ-equivalent state $\ket{\psi_{GHZ}}$ can be
  SLOCC transformed into $\ket{\phi}$
  iff $\rk(\psi_{GHZ}) \geq \rk(\phi)$.
  \qed
\end{observation}

Two related types of phenomena studied in entanglement theory are multi-copy and entanglement-assisted entanglement transformations.
Given a source state $\ket{\psi}$ and a target state $\ket{\phi}$, if
there is an integer $k$ such that the transformation of $\ket{\psi}^{\otimes k}$
to $\ket{\phi}^{\otimes k}$ can be achieved by LOCC, then we say that
$\ket{\psi}$ can be transformed to $\ket{\phi}$ by
\emph{multiple-copy entanglement transformation (MLOCC)}.
Similarly, if there is a state $\ket{c}$ such that the transformation of
$\ket{\psi}\otimes\ket{c}$ to $\ket{\phi}\otimes\ket{c}$  is possible by LOCC,
then we say that $\ket{\psi}$ can be transformed to $\ket{\phi}$ by
\emph{entanglement-assisted (or catalytic) transformation (ELOCC)}.
The state $\ket{c}$ is called a \emph{catalyst} for the transformation.
For bipartite pure states, it is known that both MLOCC and ELOCC are
strictly more powerful than ordinary LOCC \cite{JP99,BRS02}.
In the stochastic versions of multiple-copy
and entanglement-assisted transformations (SMLOCC
and SELOCC, respectively) we are only concerned with
non-vanishing success probability. For bipartite pure states,
a transformation is realizable by SLOCC if and only if it is possible by
SMLOCC or SELOCC, because of the multiplicativity of the Schmidt rank: $\Sch(\Psi\otimes \Phi)=\Sch(\Psi)\Sch(\Phi)$. Thus, there is no stochastic entanglement catalysis in bipartite systems.


In this Letter, we advance both topics of multipartite tensor ranks and SMLOCC/SELOCC transformations while demonstrating  how results of the first have unexpected consequences for the second.  As we show, since tensor rank is not multiplicative, there exist instances when the use of multiple copies or a catalyst can increase the conversion probability of some transformation from zero to positive.  In the first part of the paper, we describe the correspondence between homogeneous polynomials and symmetric states, and use it to bound the tensor rank of various multipartite symmetric states.
In the second part, we derive some general properties of SMLOCC and SELOCC transformations and then use results from the first part to demonstrate the feasibility of certain SMLOCC and SELOOC transformations when their corresponding SLOCC conversions are impossible.

\medskip
\noindent
\textit{Homogeneous polynomials and symmetric states.}
A symmetric multipartite state is one that is invariant under any permutation of the parties, $\ket{W_3}$ provides a tripartite example.  For such a state $\ket{\psi}$,
we can ask not only about its
tensor rank, but also about its \emph{symmetric tensor rank} $\sr(\psi)$: the smallest number of symmetric product states $\{\ket{\phi_i}^{\ox n}\}_{i=1,...,\sr(\psi)}$ to provide an expansion
$\ket{\psi} = \sum_{i=1}^{\sr(\psi)} \ket{\phi_i}^{\otimes n}$.
To estimate $\sr(\psi)$ (and thus $\rk(\psi)$)
we introduce a correspondence between symmetric states and homogeneous polynomials.



A homogeneous polynomial $h$ of order $N$ in $d$ variables
$x_1,\ldots,x_d$ is a linear combination of monomials
$x^{\underline{j}} = x_1^{j_1} \cdots x_d^{j_d}$ (with a multi-index
$\underline{j} = j_1\ldots j_d$), i.e.~it has the form
\(
  \displaystyle
  h = h(x_1, \cdots, x_d)
    = \sum_{\underline{j}=j_1,\ldots,j_n} a_{\underline{j}} \prod_{i=1}^d x_i^{j_i},
\)
where the sum extends over all multi-indices with $\sum_{i=1}^d j_i = N$.
Every homogeneous polynomial has a symmetric decomposition
$h = \sum_{i=1}^{\pr(h)} (\b_{1,i} x_1 + \ldots + \b_{n,i} x_d)^N$,
with the minimum number $\pr(h)$ of power terms.
We refer to this number as the \emph{polynomial rank} of $h$.
The computation and estimation of polynomial rank is a much-studied
problem in algebraic geometry~\cite{cgl08,lt09}.

Now, introducing a computational
basis $\{ \ket{1},\ldots,\ket{d} \}$ of the $d$-dimensional
local systems $\h_\alpha$, a monomial $x^{\underline{j}}$ is
associated with the Dicke state defined as
\begin{equation*}
  \ket{D(\underline{j})} \! := \! {N \choose j_1 \ldots j_d}^{\!1/2} \!\!
               P_{\textrm{sym}}
               \bigl( \ket{1}^{\otimes j_1} \otimes \cdots \otimes \ket{d}^{\otimes j_d} \bigr),
\end{equation*}
where $P_{\textrm{sym}}$ is the projection onto the Bosonic (fully symmetric)
subspace, $P_{\textrm{sym}} = \frac{1}{N!} \sum_{\pi\in S_N} U_\pi$, the
sum extending over all permutation operators $U_\pi$ of the $N$ systems.
General homogeneous polynomials (symmetric states) are associated by
linear extension of the above since monomials (Dicke states) form a basis for the homogeneous polynomials (symmetric states). That is:

\begin{observation}
  \label{le:symmetrichomogeneous}
  Every symmetric state $\ket{\psi} \in (\mathbb{C}^d)^{\otimes N}$
  is uniquely associated with a homogeneous polynomial $h(\psi)$ of
  order $N$ in $d$ variables, and vice versa each homogeneous polynomial
  $h$ is associated with a symmetric state $\ket{h}$, such
  that $h(D(\underline{j})) = x^{\underline{j}}$ and
  $\ket{x^{\underline{j}}} = \ket{D(\underline{j})}$.  Under this identification,
  symmetric tensor rank and polynomial rank are identical:
  $\pr(h) = \sr(h)$.
  \qed
\end{observation}

E.g., two copies of $\ket{W_3}$ read $\ket{W_3}^{\ox 2}=(\ket{003}+\ket{030}+\ket{300})+(\ket{012}+\ket{021}+\ket{102}+\ket{120}+\ket{201}+\ket{210})$, which is a sum of two Dicke states having corresponding homogenous polynomials $x_0x_0x_3$ and $x_0x_1x_2$. These have symmetric expansions $x_0x_0x_3=\tfrac{1}{6}((x_0+x_3)^3-(x_0-x_3)^3-2x_3^3)$ and $x_0x_1x_2=\tfrac{1}{24}((x_0+x_1+x_2)^3-(-x_0+x_1+x_2)^3-(x_0-x_1+x_2)^3-(x_0+x_1-x_2)^3)$, thus $\rk(W_3^{\ox 2}) \leq \sr(W_3^{\ox 2}) \leq 7$, which is tight~\cite{ycg09}.

Using Observation~\ref{le:ghztype},
we prove the following relations between unrestricted and symmetric tensor ranks.
\begin{theorem}
  {\upshape (a)} For multiqubit Dicke states $\ket{D(m,n)} := P_{sym}(\ket{0^{\ox m}, 1^{\ox n}})$
  with $m\geq n$, $\rk(D(m,n))=\sr(D(m,n))=m+1$,\\ {\upshape (b)} for any $N$-partite symmetric state
  $\ket{\psi}$, $\rk(\psi)\leq\sr(\psi)\leq 2^{N-1}\rk(\psi)$, \\{\upshape (c)}
  $\lim_{n \rightarrow \infty} \sqrt[n]{ \sr(\psi^{\ox n}) }= \lim_{n \rightarrow \infty} \sqrt[n]{ \rk(\psi^{\ox n}) }$.
\end{theorem}
\begin{proof}
  (a) The second equality follows from \cite[Cor.~4.5]{lt09} and
  it always holds that $\mathrm{rk}(D(m,n)) \leq
  \mathrm{srk}(D(m,n))$. So to prove the first equality, it suffices to show that the lower
  bound of $\mathrm{rk}(D(m,n))$ equals $m+1$ too.
  We use induction on $n$. For $n=1$, the claim is true~\cite{DVC00}, and we assume it holds for $n-1$.
  Ignoring normalization, we can rewrite the state as
  $\ket{D(m,n)} = \ket{D(m,n-2)}  \ket{11}
  + \ket{D(m-1,n-1)} (\ket{01} + \ket{10}) + \ket{D(m-2,n)}  \ket{00}.
  $
  Now we perform the global operation $\ket{1}\bra{11}+\frac12\ket{0}(\bra{01}+\bra{10})$
  on the last two systems which cannot increase the rank.  The resulting $(m+n-1)$-partite state is
  just the Dicke state $\ket{D(m,n-1)}$ and so $\mathrm{rk}(D(m,n)) \geq
  \mathrm{rk}(D(m,n-1)) = m+1$.

  (b)  Suppose that $\ket{\psi}$ has an optimal product state expansion $\sum_{i=1}^{\rk(\psi)}\ket{A_i}\otimes...\otimes\ket{N_i}$.
  As $\ket{\psi}$ is symmetric, we have $\ket{\psi}=\sum_{i=1}^{\rk(\psi)}P_{sym}\left (\ket{A_i}\otimes...\otimes\ket{N_i}\right )$.
  But this is just a sum of $\rk(\psi)$ Dicke states, each one corresponding to the monomial $x_{A_i}....x_{N_i}$.
  From \cite[Prop.~11.6]{lt09}, $\pr(x_{A_i}....x_{N_i})\leq 2^{N-1}$ which proves the claim.

  Part (c) follows directly from (b).
\end{proof}

%
%
%
\medskip
\noindent\textit{Three copies of $\ket{W_3}$.}
By Observation~\ref{le:symmetrichomogeneous}, the homogeneous
polynomial $h(W_3^{\ox 3})$ can be written as
$\frac{2}{9}(x_0 x_1 x_6 + x_0 x_2 x_5 + x_0 x_3 x_4 + x_1 x_2 x_4)+\frac{1}{9} x_0^2 x_7$.
To compute its polynomial rank, we perform the
following linear transformations which do not change the polynomial
rank: $y_1=x_1+x_2-x_4, y_2=x_1-x_2+x_4, y_4=-x_1+x_2+x_4, z_3=1/2(x_3+x_5), z_5=1/2(x_3+x_6), z_6=1/2(x_5+x_6)$. By using
the fact that the polynomial rank is invariant under scalar
multiplication, we can remove constant coefficients and obtain
  $\pr(h(W_3^{\ox 3}))
    \leq \pr( x_0 y_1 z_6 - y_1^3 )
            +
          \pr( x_0 y_2 z_5 - y_2^3 )
            +
          \pr( x_0 y_4 z_3 - y_4^3 )
            +
          \pr\bigl( (y_1+y_2+y_4)^3 + x_0^2 x_7 \bigr)
     \leq 16.$
Here, the
inequalities follow from~\cite[Table 2]{lt09}. With the lower bound
$\rk(W_3^{\ox 3}) \geq 15$ \cite{ycg09}, we have
\begin{theorem}
  \label{thm:3copy3W}
  {\upshape (a)} $\rk(W_3^{\ox 3}) = 15 \text{ or } 16$, \\{\upshape(b)} $\lim_{n\rightarrow \infty}
  \sqrt[n]{\rk(W_3^{\ox n})}
  \leq
  \sqrt[3]{16} \approx 2.52$.
  \qed
\end{theorem}
This improves the previously best bound of
$\rk(W_3^{\ox 3}) \leq 21$~\cite{ycg09}. In particular,
Theorem~\ref{thm:3copy3W} implies that two tripartite GHZ-type states with
tensor rank 4 are sufficient to prepare three $\ket{W_3}$ states
under SLOCC.
%

\medskip
\noindent\textit{Upper bound on the tensor rank of
$\ket{W_N}^{\otimes n}$.}
The $N$-partite W state is defined as the
Dicke state $\ket{W_N} = \frac{1}{\sqrt{N}}(\ket{0\ldots 01} +
\ldots + \ket{10\ldots 0})\in (\mathbb{C}^2)^{\otimes N}$.  As
$\ket{W_N}^{\otimes n}$ will be a linear combination of Dicke
states, we can obtain an upper bound for $\rk(W_N^{\otimes n})$ by
adding up the tensor ranks of each component Dicke state.  Now each
one of these corresponds exactly to a different way of separating
$n$ distinct excitations $\ket{1}$ into $k = 1, 2, \ldots, N$ local
states.  This number is equal to the Stirling number of the second
kind, namely $S(n,k) = \frac{1}{k!} \sum_{i=0}^k
(-1)^{k-i} {k \choose i} i^n$, where we only have to consider $k
\leq n$ as any larger number of parties is taken care of by the
symmetrization.

For example, $S(3,2)=3$ which implies  that there are three ways of
separating three excitations into two local systems, namely
$\ket{0^{\ox 3},\ldots,0^{\ox 3},100,011}$, $\ket{0^{\ox
3},\ldots,0^{\ox 3},010,101}$ and $\ket{0^{\ox 3},\ldots,0^{\ox
3},001,110}$. By permuting the local states, each of these generates
a Dicke state with corresponding monomials $x_0^{N-2} x_4 x_3$,
$x_0^{N-2} x_2 x_5$ and $x_0^{N-2} x_1 x_6$, respectively.

Since each of the $S(n,k)$ monomials representing the same
separation ($n \rightarrow k$) are related by a simple change in
variables, each will have the same polynomial rank. Then by adding
up all separations we obtain
$  \rk(W_N^{\ox n})
\leq \sum^{\min\{N,n\}}_{k=1}
                                      S(n,k) \pr(x_0^{N-k} x_1 \cdots x_k)
                         \leq \sum^{\min\{N,n\}}_{k=1}
                                      S(n,k) (1 \!+\! \max\{ N \!-\! k, k \}) 2^{k-1},
$
where the second inequality follows from~\cite[Cor.~4.5
and Prop.~11.6]{lt09}. In particular, this bound is of the
form $f(n) N + g(n)$ with some functions $f(n)$ and $g(n)$.
In other words,
\begin{theorem}
  \label{thm:ncopyNW<N/2}
  $\rk(W_N^{\ox n})$ is upper bounded by a linear
  function in $N$.
  Thus for large $N$, $\ket{W_N}^{\ox n}$ can be prepared
  by LOCC from a GHZ-type state of rank linear in $N$.
  \qed
\end{theorem}

The large-$n$ behavior of this bound is not very good, but based
on a simple asymptotic consideration of the Stirling numbers
for $n \approx \log N$, we find that
\begin{corollary}
  \label{cor:rrk-W_N}
  $\lim_{n\rightarrow \infty} \sqrt[n]{\rk(W_N^{\ox n})} \leq O(\log N)$.
  \qed
\end{corollary}

\noindent\textit{Lower bound on the tensor rank of $\ket{W_N}^{\otimes n}$.}
\begin{lemma}
\label{le:equivalentsloccrk}
Any state of the form
$ \ket{\O} = \ket{W_{N-1}}^{\ox n}+\sum_{k=1}^n\sum_{\pi\in
S_N}c_{\pi k}U_{\pi}\left( \ket{W_{N-1}}^{\ox
k}\ket{0_{N-1}}^{\ox(n-k)}\right) $
is SLOCC equivalent to $\ket{W_{N-1}}^{\ox n}$.
\end{lemma}
\begin{proof}
We perform successively invertible SLOCC transformations on
$\ket{\O}$,
each transformation eliminating a term in the double sum.
For instance, applying the transformation
$\ket{W_{N-1}}\to\ket{W_{N-1}}-c_{\pi k}\ket{0_{N-1}}$,
$\ket{0_{N-1}}\to \ket{0_{N-1}}$ on $\ket{\O}$ by local invertible operators
will eliminate the term $U_{\pi}\left( \ket{W_{N-1}}^{\ox
k}\ket{0_{N-1}}^{\ox(n-k)}\right)$.
The procedure is repeated on all terms in the sum until just
$\ket{W_{N-1}}^{\ox n}$ remains.
\end{proof}

To prove a lower bound, note that
$\rk(W_N^{\ox n})$ is the minimum number of product states whose
linear span contains the set $S=\{\ket{W_{N-1}},\ket{0_{N-1}}\}^{\ox
n}$.  Each of these product states can be substituted with an element from $S\setminus\{\ket{W_{N-1}}^{\ox n}\}$ to yield a new set whose linear span also contains $S$.  Thus, $\ket{W_{N-1}}^{\ox n}$ is a linear combination of elements from  $S\setminus\{\ket{W_{N-1}}^{\ox n}\}$ and at most $\rk(W_N^{\ox n})-(2^n-1)$ product states.  Thus by Lemma~\ref{le:equivalentsloccrk} we get $\rk(W_{N-1}^{\ox n}) \leq \rk(W_{N}^{\ox n})
- (2^n-1)$.
As proven
in~\cite{ycg09}, for $N=3$, $2^{n+1}-1\leq\rk(W_3^{\ox n})$.  From
these two inequalities, a simple inductive argument provides part
(a) in the next theorem; part (b) then immediately follows after
observing that Theorem~\ref{thm:ncopyNW<N/2} reads
$\rk(\ket{W_{N}}^{\ox 2}) \leq 3N-2$ when $n=2$.
\begin{theorem}
  \label{thm:ncopyNWlowerbound}
  {\upshape (a)} $\rk(W_{N}^{\ox n}) \geq (N-1) 2^n -N+2$, \\ {\upshape (b)} $\rk(W_{N}^{\ox 2}) = 3N-2$.
  \qed
\end{theorem}

\noindent\textit{Multi-copy and catalytic SLOCC transformations.} We
now move on to the topic of SLOCC catalysis for multipartite
entanglement transformations.  Let $\h=\bigotimes_{k=1}^n\h_k$ and
$\h'=\bigotimes_{k=1}^n\h_k'$ be $n$-partite quantum systems, and
consider $\h_k$ and $\h_k'$ to be orthogonal to each other. Let
$\ket{\psi_0}$ and $\ket{\psi_1}$ be two vectors from $\h$ and
$\h'$, respectively. Then the direct sum of $\ket{\psi_0}$ and
$\ket{\phi_1}$ is given by
 $\ket{\psi_0}\oplus \ket{\psi_1} \in \h\oplus\h'\subseteq
\bigotimes_{k=1}^n (\h_k\oplus\h_k')$. Notice that when
$\ket{\phi_1}=\otimes_{k=1}^n L_k \ket{\psi_1}$ and
$\ket{\phi_2}=\otimes_{k=1}^n L_k' \ket{\psi_2}$, we simply have $
\ket{\phi_1}\oplus \ket{\phi_2} = \bigotimes_{k=1}^n (L_k\oplus
L_k')(\ket{\psi_1}\oplus \ket{\psi_2})$. By induction one can
immediately show that the SLOCC ordering is preserved under direct
sums.
\begin{lemma}\upshape\label{le:directsum}
If $\ket{\psi_k}$ can be transformed into $\ket{\phi_k}$ via SLOCC,
then $\bigoplus_{k} \ket{\psi_k}$ can also be transformed
into $\bigoplus_{k}\ket{\phi_k}$ via SLOCC.
\qed
\end{lemma}

We can use Lemma~\ref{le:directsum} to get a general relation
between SMLOCC and SELOCC. Assume that $\ket{\psi}^{\otimes n}$ can
be transformed into $\ket{\phi}^{\otimes n}$ via SLOCC for some
$n\geq 1$. Then by choosing $ \ket{c}=\bigoplus_{k=1}^{n}
\ket{\psi}^{\otimes n-k}\otimes \ket{\phi}^{\otimes k}$, the result
that $\ket{\psi}\otimes\ket{c}$ can be transformed to
$\ket{\phi}\otimes\ket{c}$ via SLOCC follows from
Lemma~\ref{le:directsum}. So we get, similar to \cite{DFLY05,DFY05}:
\begin{theorem}
\label{thm: SMLOCC->SELOCC}
  If $\ket{\psi}$ can be transformed to $\ket{\phi}$ via SMLOCC,
  then the same transformation can also be achieved via SELOCC.
  \qed
\end{theorem}


By Observation~\ref{le:ghztype}, to demonstrate the effect of
entanglement catalysis, we only need to find a state $\ket{\phi}$
with the following property: $\rk(\phi)=n$ and there is some $k\geq
1$ such that $\rk(\phi^{\otimes k})\leq (n-1)^k$. The source state
$\ket{\psi}$ can be chosen as an $n$-partite GHZ state  with tensor
rank $(n-1)$.  Such states $\ket{\phi}$ do exist as proven in the
previous section. In the following we shall provide two different
constructions. The first class is given by the famous tripartite
matrix multiplication tensor and the second one is given by the
$W_N$ states. By Theorem~\ref{thm: SMLOCC->SELOCC} these also
suffice to show the existence of SELOCC transformations when the
uncatalyzed transformation is impossible.

\begin{theorem}
\label{matrix_tensor}
Let  $\ket{\Phi^{(3)}}=\ket{\Phi_2}_{AB}\otimes \ket{\Phi_2}_{BC}\otimes\ket{\Phi_2}_{CA}$,
where $\ket{\Phi_2}=\ket{00}+\ket{11}$, and let
$\ket{\psi}_{ABC}$ be any generalized GHZ-type state with tensor rank $6$.
Then the transformation of $\ket{\psi}$ to  $\ket{\Phi^{(3)}}$
cannot be realized by SLOCC but can be realized by both SMLOCC and SELOCC.
\end{theorem}
\begin{proof}
It has been shown that $\ket{\Phi^{(3)}}$ is just the $2\times 2$
matrix multiplication tensor~\cite{Strassen-1969a, CDS08}.
By a well known result in algebraic complexity theory, $\rk(\Phi^{(3)})=7>6$~\cite{Winograd-1971a}.
Hence, $\ket{\psi}$ cannot be SLOCC transformed into $\ket{\Phi^{(3)}}$.
Now the best known algorithm for $d\times d$ matrix multiplication
requires $O(d^{2.376})$ multiplication steps \cite{CW90}. Hence the
tensor rank of $\ket{\Phi^{(3)}}^{\otimes n}$, which corresponds to
the algebraic complexity of $2^n\times 2^n$ matrix multiplication,
is $O(2^{2.376n})$.
On the other hand, the tensor rank of $\ket{\Psi}^{\otimes n}$ is
simply $6^n=2^{(\log_2 6) n} \approx 2^{2.585 n}$, which is larger
than $O(2^{2.376n})$ for sufficiently large $n$.
Thus we have confirmed the existence of $n$ (perhaps very large) such that
  $\rk(\Psi^{\otimes n})\geq \rk((\Phi^{(3)})^{\otimes n})$.
Hence both SMLOCC and SELOCC are possible.
\end{proof}

If we consider multipartite rather than tripartite state spaces,
the $W$ states provide much simpler examples.

\begin{theorem}
  \label{multipartite_W}
  For any $N\geq 5$ the transformation of $\ket{\GHZ_N^{N-1}}$ to $\ket{W_N}$
  cannot be realized by SLOCC but can be achieved by both SMLOCC and SELOCC.
  Furthermore, two copies are sufficient in SMLOCC, and the catalyst in
  SELOCC can be chosen as $\ket{W_N}\oplus\ket{\GHZ_N^{N-1}}$.
\end{theorem}
\begin{proof}
The result follows immediately from the facts that
$\rk(W_N^{\otimes 2}) = 3N-2$ in Theorem~\ref{thm:ncopyNWlowerbound} and
$(N-1)^2\geq 3N-2$ for $N\geq 5$.
One can easily see that the rank of the GHZ state can indeed
be chosen as $\lceil\sqrt{3N-2}\rceil$,
which is much smaller than $N-1$ for $N \gg 1$.
\end{proof}

\medskip
\noindent\textit{Conclusions.}
We have shown that the theory of homogeneous polynomials can be
used to obtain insights on the symmetric tensor rank of symmetric states.  Via this connection, we proved upper and lower bounds on the tensor rank for one and multiple copies of $W_N$ states as well as the exact tensor and symmetric tensor rank of multiqubit Dicke states.  We then proceeded to show that multi-copy and catalytic activation
of otherwise impossible SLOCC transformations exists, using our
results on $W_N$ states to find explicit low-dimensional examples.

Our work suggests several open questions which we leave
for future investigation.  First, given two states $\ket{\psi}$ and $\ket{\phi}$, what are the necessary and sufficient conditions such that $\ket{\psi}$ can be converted to $\ket{\phi}$ under SMLOCC and SELOCC?  When $\ket{\psi}$ is a generalized GHZ state, the question becomes completely a matter of tensor rank multiplicativity.  Asked in a different way, for some target state $\ket{\phi}$, when does there exist a state $\ket{\psi}$ such that transformation $\ket{\psi}$ to $\ket{\phi}$ is possible under SMLOCC and SELOCC but impossible with just single copies.
Another relevant problem is to determine the asymptotic tensor rank of $\ket{W_3}$, and
more generally of $\ket{W_N}$. Note that our lower bound of $2$
coincides with the \emph{border rank} \cite{strassen}. It is conceivable
that the asymptotic rank is $2$ for all $N$, but even an
improvement of our logarithmic upper bound would be interesting.




\medskip

\acknowledgments

We thank Dr. Nengkun Yu for helpful discussion over the fact that
tensor rank of $\ket{W_N}^{\ox n}$ is linear in $N$. EC is partially
supported by the U.S. NSF under Awards 0347078 and 0622033. RD is
partly supported by QCIS, University of Technology, Sydney, and the
NSF of China (Grant Nos.~60736011 and 60702080). AW is supported by
the E.C., the U.K.~EPSRC, the Royal Society and a Philip Leverhulme
Prize. The CQT is funded by the Singapore MoE and the NRF as part of
the Research Centres of Excellence programme. Research at Perimeter
Institute is supported by the Government of Canada through Industry
Canada and by the Province of Ontario through the Ministry of
Research \& Innovation.

\end{document}